\documentclass{article}

\usepackage{arxiv}

\usepackage[utf8]{inputenc} 
\usepackage[T1]{fontenc}    
\usepackage{url}            
\usepackage{booktabs}       
\usepackage{amsfonts}       
\usepackage{nicefrac}       
\usepackage{microtype}      
\usepackage{graphicx}
\usepackage{doi}
\usepackage[all]{xy}
\usepackage{setspace}
\usepackage{xcolor}
\usepackage{tikz}
\usetikzlibrary{cd}

\title{Comparison of some geometric frameworks for dissipative evolution in multiscale non-equilibrium thermodynamics}

\author{Miroslav Grmela\\
\'{E}cole Polytechnique de Montr\'{e}al,
  C.P.6079 suc. Centre-ville,\\
 Montr\'{e}al, H3C 3A7,  Qu\'{e}bec, Canada\\
\And Michal Pavelka \\
 Mathematical Institute, Faculty of Mathematics and Physics, Charles University,\\
 Sokolovsk\'{a} 83, 18675 Prague, Czech Republic\\
Corresponding author: pavelka@karlin.mff.cuni.cz
}

\newcommand{\shorttitle}{Geometric dissipation}

\usepackage{amsmath,amssymb,amsfonts,latexsym,float,graphics,epsfig}

\newcommand{\qq}{\mathbf{q}}
\newcommand{\rr}{\mathbf{r}}
\newcommand{\pp}{\mathbf{p}}
\newcommand{\mm}{\mathbf{m}}
\newcommand{\oomega}{\boldsymbol{\omega}}
\newcommand{\ww}{\mathbf{w}}
\newcommand{\xx}{\mathbf{x}}
\newcommand{\FF}{\mathbf{F}}
\newcommand{\yy}{\mathbf{y}}
\newcommand{\vv}{\mathbf{v}}
\newcommand{\MM}{\mathbf{M}}
\newcommand{\JJ}{\mathbf{J}}
\newcommand{\LL}{\mathbf{L}}
\newcommand{\RRR}{\mathfrak{R}}
\newcommand{\rrho}{\mathfrak{r}}
\newcommand{\tqqs}{\tilde{\mathbf{q}}^*}
\newcommand{\gtqq}{\widetilde{\nabla\mathbf{q}^*}}
\newcommand{\tss}{\tilde{s}^*}
\newcommand{\gtss}{\widetilde{\nabla s^*}}
\newcommand{\gmetric}{\mathbf{g}}
\newcommand{\XX}{\mathbf{X}}
\newcommand{\YY}{\mathbf{Y}}
\newcommand{\KK}{\mathcal{K}}

\begin{document}
\maketitle

\begin{abstract}
In this paper, we review and compare some geometric frameworks for dissipation in non-equilibrium thermodynamics. We start with a brief overview of classical irreversible thermodynamics and gradient dynamics. Then we discuss several specific frameworks including Rayleigh dissipation potential and the dissipative d'Alembert framework, showing their relations with gradient dynamics. Finally, we discuss frameworks for dissipative evolution generated from Poisson brackets.
\end{abstract}

\tableofcontents

\section{Introduction}

Dissipation terms in vector fields of mesoscopic dynamical theories bring about the approach of the trajectories generated by the vector fields to trajectories generated by vector fields of other mesoscopic dynamical theories (which we shall call \textit{target theories}) that involve fewer microscopic details. In particular, the target theory can be classical equilibrium thermodynamics. In such a case, there is no time evolution in the target theory; the target trajectories are fixed points of the original vector field. In another example, we can consider fluid mechanics as a target theory and kinetic theory as the initial mesoscopic theory.
The approach (which is brought about by the dissipation), in which unimportant microscopic details are being erased and overall features in the phase portrait are emerging, is expressed mathematically in the time evolution of \textit{entropy}, which is a potential of nonmechanical origin addressing overall features in phase portraits and playing the role of a Lyapunov function.

Historically, the first passage from the microscopic particle dynamics to classical equilibrium thermodynamics was made by Ludwig Boltzmann \cite{boltzmann} for an ideal gas. The Boltzmann entropy introduced in the passage indeed plays the role of a Lyapunov function for the approach to thermodynamic equilibrium states, and if the entropy is evaluated at the approached equilibrium states, it becomes the fundamental equilibrium thermodynamic relation for the ideal gas. In the Gibbs passage \cite{gibbscw} from the microscopic particle dynamics to classical equilibrium thermodynamics (which applies to all macroscopic systems), the time evolution, in the course of which the Gibbs entropy is maximized, is absent. Many frameworks have been developed for the passage from fluid mechanics to classical equilibrium thermodynamics. Fluid mechanics (classical and extended) is supplemented with the assumption of local equilibrium (classical and extended) and the requirement that the entropy plays the role of a Lyapunov function in the passage (second law of thermodynamics). 

Technically speaking, if we do not restrict the stability analysis to the reachable submanifold of the whole space of state variables (states with the same energy, momentum, etc., as the equilibrium state), the entropy is rather an S-function \cite{gpb1986} than a Lyapunov function \cite{lyapunov}. However, for simplicity, we shall call it a Lyapunov function, tacitly assuming the restriction to the reachable submanifold, which is often reflected in the choice of vector spaces in the case of infinite-dimensional systems \cite{vitek-entropy}. 

Another technical issue that we are not addressing in this paper is the enhancement of dissipation by the presence of a nondissipative Hamiltonian term in the vector field. In the context of the Boltzmann equation, the vector field without the Hamiltonian term drives trajectories to the local Maxwell distribution, with the Hamiltonian term to the total Maxwell distribution (\cite{desvilettes-villani}). We  conjecture that this type of enhancement of dissipation plays an essential role in the emergence of dissipation in dynamics of macroscopic systems. Ignorance of details of
a minute irregularity that may arise on microscopic scale is enhanced  to the dissipation detectable on the  macroscopic scale.

Note also that in the case of infinite-dimensional systems, our calculations are only formal, since we do not specify the function spaces and the smoothness of the functionals. In particular, we do not specify the functional spaces for the state variables, the entropy, the energy, and the dissipation potential. We assume that all functionals are sufficiently smooth (twice Fréchet differentiable) and local (with vanishing boundary conditions).

Among the frameworks, we mention Classical Irreversible Thermodynamics (CIT) \cite{onsager1930,onsager1931,dgm,prigogine-tip}, Extended Irreversible Thermodynamics \cite{jou-eit,lebon-understanding}, Internal Variables Thermodynamics \cite{van-berezovski}, Rational Thermodynamics \cite{trus}, Rational Extended Thermodynamics \cite{mr}, Steepest Entropy Ascent \cite{sea}, and port-Hamiltonian systems \cite{schaft-survey,maschke-schaft}.

In this paper, we focus on geometric frameworks for multiscale dissipative time evolution. Such frameworks are formulated in the multiscale setting (i.e., they can be used in the context of fluid mechanics, kinetic theory, and other autonomous mesoscopic theories) and provide a unification and additional physical and mathematical insights.
Among the geometric frameworks, we discuss generalized gradient dynamics \cite{otto,gyarmati}, the General Equation for Non-Equilibrium Reversible-Irreversible Coupling (GENERIC) \cite{grcontmath,grpla84,go,og,hco,pkg,be}, and metriplectic systems \cite{mor,kauf,grcontmath}, where gradient dynamics is combined with Hamiltonian mechanics. In the case when the target mesoscopic theory is not equilibrium thermodynamics but a mesoscopic theory involving fewer details, the role of the entropy is replaced by the Rayleigh dissipation potential \cite{Rayleigh,landau5,dv}. Dissipative terms can be also added as constraints in the variational formulation of mechanics, leading to the d'Alembert variational principle in non-equilibrium thermodynamics \cite{fgb19-variational,yoshimura-hamiltonian,fgb19-variational-thermodynamics}. Finally, when we have a mechanical (Hamiltonian) system, dissipative evolution can be constructed from the underlying Poisson bracket and from the Hamiltonian (double bracket dissipation \cite{bloch-krishnaprasad-marsden1996} and Ehrenfest regularization \cite{ehre}).

In particular, we show that the Rayleigh-potential framework can be cast into the GENERIC form, employing generalized gradient dynamics as the dissipative part of the evolution. We also formulate a version of the dissipative d'Alembert variational principle using the Rayleigh dissipation potential, which makes it easier to eliminate fast-relaxing variables in a geometric manner. Finally, we show a variational formulation of the GENERIC framework and a relation of Extended GENERIC with port-Hamiltonian systems \cite{schaft-survey}. 

When comparing these frameworks, we focus on the following features: (i) Geometric formulation, since frameworks that are not geometrically consistent may not be invariant with respect to changes of variables; (ii) Modeling minimalism, since approaches that require fewer modeling assumptions are preferable; (iii) Generality, so that framework can be used on various levels of description.

In Section \ref{sec.cit}, we motivate the frameworks for dissipative evolution by pointing out some problems of CIT. Section \ref{sec.gd} contains generalized gradient dynamics while Section \ref{sec.ray} is about the Rayleigh dissipation potential. In Section \ref{sec.var} we discuss the dissipative d'Alembert variational principle, and in Section \ref{sec.ham} we recall some frameworks where dissipative evolution is constructed from the Poisson bracket. Sections \ref{sec.ph} and \ref{sec.contact} show the connection of GENERIC with port-Hamiltonian systems and contact geometry, respectively.

\section{Classical irreversible thermodynamics}\label{sec.cit}
State variables in the context of Classical Irreversible Thermodynamics \cite{dgm} are chosen to be $\xx = (\yy,e)$, where $e$ is the volumetric total energy density and $\yy$ represents the remaining densities that play the role of state variables (for instance mass density, momentum density, or some internal variable densities). The time evolution of the volumetric energy density $e$ is governed by
\begin{equation}
    \partial_t e = -\nabla\cdot \JJ_e,
\end{equation}
where $\JJ_e=e\vv + \JJ_q$ is the total energy flux, $\vv$ is the barycentric velocity and $\JJ_q$ is the heat flux (see \cite{dgm} for other definitions of heat flux). This evolution equation expresses the conservation of the total energy. 

The evolution equations of the remaining state variables $\yy$ are expected in the form 
\begin{equation}
    \partial_t \yy = -\nabla\cdot(\yy \otimes \vv + \JJ_\yy) + \sigma_\yy,
\end{equation}
where $\JJ_\yy$ is the flux of $\yy$ and $\sigma_\yy$ is a source term.

The unknown fluxes and source terms are chosen so that they fulfill the second law of thermodynamics and Onsager-Casimir reciprocal relations. This is achieved by writing the balance equation for the volumetric entropy density $s(\xx)$ and by rewriting the entropy production as a sum of products of thermodynamic fluxes and forces. The entropy balance equation then reads
\begin{align}
    \partial_t s(\xx) =& \frac{\partial s}{\partial \xx}\cdot \partial_t \xx
    = \frac{\partial s}{\partial e}\left(-\nabla\cdot \JJ_e\right) + \frac{\partial s}{\partial \yy}\cdot \partial_t \yy\nonumber\\
    =& -\nabla\cdot\left(\vv e \frac{\partial s}{\partial e} + \JJ_q \frac{\partial s}{\partial e} + \vv \left(\yy\cdot\frac{\partial s}{\partial \yy}\right) + \JJ_\yy \frac{\partial s}{\partial \yy}\right) + \JJ_e\cdot \nabla \frac{\partial s}{\partial e} + (\vv\cdot\nabla)\frac{\partial s}{\partial \yy}\cdot \yy +\JJ_\yy \cdot \nabla \frac{\partial s}{\partial \yy} + \sigma_\yy \cdot \frac{\partial s}{\partial \yy}\nonumber\\
    =& -\nabla\cdot\left(\frac{\partial s}{\partial e}\JJ'_q + \left(e\frac{\partial s}{\partial e} + \yy\cdot\frac{\partial s}{\partial \yy}\right)\vv\right)
    + \left( e \nabla \frac{\partial s}{\partial e} + \yy \cdot \nabla \frac{\partial s}{\partial \yy}\right)\vv
    + \JJ'_q \cdot \nabla \frac{\partial s}{\partial e} + \JJ_\yy \cdot \nabla \frac{\partial s}{\partial \yy} + \sigma_\yy \cdot \frac{\partial s}{\partial \yy}
\end{align}
where the derivative of entropy with respect to energy is the inverse temperature, $T=\left(\frac{\partial s}{\partial e}\right)^{-1}$ is the temperature, and the reduced heat flux is defined as $\JJ'_q = \JJ_q- \JJ_\yy\cdot\left(\frac{\partial e}{\partial \yy}\right)_s$.

At this point, we need to use the Gibbs-Duhem relation (or extensivity of entropy in terms of the densities), which states that
\begin{equation}
    e \nabla \frac{\partial s}{\partial e} + \yy \cdot \nabla \frac{\partial s}{\partial \yy} = 0
   \quad\text{and}\quad
    e \frac{\partial s}{\partial e} + \yy \cdot \frac{\partial s}{\partial \yy} = s.
\end{equation} 
Finally, when the entropy derivatives are expressed in terms of the corresponding energy derivatives, the entropy balance equation becomes
\begin{equation}
    \partial_t s = -\nabla\cdot\left(T^{-1}\JJ'_q + s\vv\right)
    + \underbrace{\frac{1}{T}\JJ'_q \cdot \left(-T^{-1}\nabla \frac{\partial e}{\partial s}\right) 
    +\JJ_\yy \cdot\left(-\nabla \frac{\partial e}{\partial \yy}\right) 
    +\sigma_\yy\cdot \left(-\frac{\partial e}{\partial \yy}\right)}_{=\sigma_s},
\end{equation}
where the entropy production $\sigma_s$ is written as a sum of products of thermodynamic fluxes and forces. The total entropy flux can be identified with the terms under the divergence as $\JJ_s = T^{-1}\JJ'_q + s\vv$.

To guarantee the non-negativity of the entropy production, the fluxes and source terms are then expressed as linear combinations of the thermodynamic forces (gradients of the derivatives of energy),
\begin{equation}
    \begin{pmatrix}
        \JJ'_q\\
        \JJ_\yy\\
        \sigma_s
    \end{pmatrix}
    = \MM(\xx)\cdot
    \begin{pmatrix}
        -T^{-1}\nabla\frac{\partial e}{\partial s} \\
        -\nabla\frac{\partial e}{\partial \yy} \\
        -\frac{\partial e}{\partial \yy}
    \end{pmatrix}
\end{equation}
where $\MM$ is a linear operator that is positive semi-definite, $\xx\cdot \MM(\xx)\cdot \xx \geq 0$ and should satisfy the Onsager-Casimir reciprocal relations.

Entropy is often quadratic in the non-conserved state variables (for instance heat flux in the Cattaneo equation \cite{catt}), which leads to Extended Irreversible Thermodynamics \cite{jou-eit} and Internal Variable Thermodynamics \cite{van-berezovski,maugin,Mongiovi2017}.

Although this construction leads to both energy conservation and non-negative entropy production, it has the following drawbacks:
\begin{itemize}
\item Evolution of state variables $\yy$ does not in general have to be in the divergence form, as in the case of polymeric fluids (conformation tensor is not conserved) \cite{bird2}, magnetohydrodynamics (where the Godunov-Powell terms are important for numerical stability \cite{god}), Reynolds stress in turbulence \cite{miroslav-turbulence}, and others \cite{pkg}. Instead, the evolution equation for $\yy$ can have a part that is in the divergence form and a part that is not. Moreover, we can move some terms between the two parts, which makes the identification of fluxes and forces ambiguous.
\item If $\MM$ is replaced with $\MM^s+\MM^a$, where $\MM^s$ is symmetric and $\MM^a$ is a skew-symmetric operator, the entropy production will not be affected. Therefore, when we construct the constitutive relation based only on the entropy production, we cannot determine what the possible skew-symmetric part $\MM^a$ looks like, as any skew-symmetric coupling would not change the entropy production. In practice, this means that CIT cannot tell which type of derivative (material, upper convected, lower convected, Lie, etc.) works for a particular field \cite{pkg}.
\item Linear force-flux relations are only valid in the vicinity of the thermodynamic equilibrium, where the thermodynamic forces are small, but in general, non-linear closures are possible and often needed \cite{grchemkin,pkg}.
\end{itemize}
Due to these problems, other frameworks for dissipation have been developed, which we review in the following sections.

\section{Gradient dynamics}\label{sec.gd}

\subsection{Generalized gradient dynamics}
Assume that a macroscopic system is described by state variables $\xx$. They could be, for instance, hydrodynamic fields, 1-particle or many-particle distribution functions, various fields describing internal structure of suspensions or macromolecules, etc.
The time evolution of $\xx$ is governed by generalized gradient dynamics generated by a dissipation potential that may be quadratic (gradient dynamics \cite{otto}) or more general \cite{grchemkin,nonconvex,mielke-peletier}. The time evolution equations of the state variables are then
\begin{equation}\label{eq.GD}
    \dot{\xx} = \frac{\delta \Xi}{\delta \xx^*}\Big|_{\xx^*=\frac{\delta S}{\delta \xx}}
\end{equation}
where $\Xi$ is a dissipation potential, $S$ is the entropy, and $\xx^*$ are thermodynamic forces conjugate to $\xx$. The derivative $\delta\bullet/\delta \bullet$ stands for the functional derivative so that we can differentiate functionals with respect to fields. The dissipation potential is a non-negative functional of $\xx^*$ that has a minimum at $\xx^*=0$, is convex in the vicinity of the minimum, and is radially monotonous, $\xx^*\cdot\frac{\delta \Xi}{\delta \xx^*}\geq 0$; see \cite{roubicek,nonconvex}. The dissipation potential $\Xi(\xx,\xx^*)$ also depends on the state variables $\xx$.

Energy conservation is guaranteed when
\begin{equation}\label{eq.Xi.deg}
    \Xi(\xx, \xx^*)\Big|_{\xx^*}
    =
    \Xi(\xx, \xx^*)\Big|_{\xx^* + \lambda \frac{\delta E}{\delta \xx}} \quad \forall \lambda \in \mathbb{R},
\end{equation}
since by taking a derivative with respect to $\lambda$ at $\lambda=0$, we obtain $\frac{\delta \Xi}{\delta \xx^*}\cdot \frac{\delta E}{\delta \xx}=0$ and consequently
\begin{equation}
    \dot{E} = \int d\rr\, \frac{\delta E}{\delta \xx}\cdot \dot{\xx} = \int d\rr\,  \frac{\delta E}{\delta \xx}\cdot \frac{\delta \Xi}{\delta \xx^*}\Big|_{\xx^*=\frac{\delta S}{\delta \xx}} = 0,
\end{equation}
see \cite{kraaj}.

For instance, in fluid dynamics, where the state variables may be chosen as the mass density, momentum density, and energy density, $\xx = (\rho,\mm,e)$, the dissipation potential is quadratic and depends on the gradients of the conjugate variables $\mm^*$ and $e^*$ (to be identified with the corresponding derivatives of entropy):
\begin{equation}
    \Xi(\xx,\xx^*) = \int d\rr\, \mu(\xx) \left(\frac{1}{2}\left(\vv\otimes \nabla e^* + \nabla e^* \otimes \vv\right) + \frac{1}{2}\left(\nabla \mm^* + (\nabla \mm^*)^T\right)\right)^2
    +\int d\rr\, \zeta(\xx)\left(\vv\cdot\nabla e^* + \nabla\cdot \mm^*\right)^2.
\end{equation}
One might expect that only the part $\frac{1}{2}(\nabla \mm^* + (\nabla \mm^*)^T)$ contributes to the dissipation, but the additional term $\frac{1}{2}(\vv\otimes \nabla e^* + \nabla e^* \otimes \vv)$ is required by the degeneracy condition \eqref{eq.Xi.deg} and Galilean invariance. The evolution equations then become
\begin{subequations}
\begin{align}
    \partial_t \mm =& \nabla\cdot\left(\mu(\nabla \vv + (\nabla \vv)^T)\right) + \nabla(2 \zeta (\nabla\cdot\vv)),\\
    \partial_t e =& -\nabla\cdot\left( \left(\mu(\nabla \vv + (\nabla \vv)^T) + 2 \zeta (\nabla\cdot\vv)\mathbf{I}\right)\cdot\vv\right),
\end{align}
\end{subequations}
where $\vv = -T\mm^* = - \frac{\delta S}{\delta \mm}/\frac{\delta S}{\delta e}$ is the velocity field and $e^* = \frac{\delta S}{\delta e}$ is the inverse temperature. When the reversible (Hamiltonian) evolution is also added (see, for instance, \cite{pkg}), the evolution equations become the compressible Navier-Stokes equations with shear and bulk viscosities $\mu$ and $\zeta$, accompanied by the continuity equation and energy conservation.

Generalized gradient dynamics also represents the most probable path of stochastic systems obeying the large-deviation principle \cite{mielke-potential,mielke-peletier,hco-jnet2020-I,hco-jnet2020-II}, so it has a foundation also in stochastic theories of mesoscopic dynamics.

Non-quadratic dissipation potentials, for instance $\Xi \propto \cosh(X)$, where $X$ is a thermodynamic force linear in $\xx^*$, are used, for instance, in the case of the Boltzmann equation or chemical kinetics \cite{grcontmath,grchemkin,pkg}.

\subsection{Gradient dynamics with constraints}
In non-equilibrium thermodynamics, the evolution is typically constrained by energy or momentum conservation. Such constraints can be incorporated into gradient dynamics \eqref{eq.GD} by means of Lagrange multipliers, which can be geometrically seen as a Morse family \cite{omg1}. 

Alternatively, the constraints $c^\alpha(\xx)$ can be incorporated into the framework by forming thermodynamic forces that are orthogonal to the gradients of the constraints \cite{hutter2013,miroslav-guide},
\begin{equation}
    X^\alpha = \KK_\alpha \xx^*
\end{equation}
where $\KK$ is a linear operator satisfying $\KK_\alpha(dc^\beta)=0 \quad\forall\alpha \forall\beta$. When the dissipation potential depends on the conjugate variables only via the thermodynamic forces, $\Xi(\xx,\xx^*) = \Xi(\xx, \KK\xx^*)$, the evolution equations become
\begin{equation}
    \dot{\xx} = \sum_\alpha \KK^\dagger_\alpha \frac{\delta \Xi}{\delta X^\alpha}\Big|_{X^\alpha = \KK_\alpha \frac{\delta S}{\delta \xx}},
\end{equation}
where $\KK^\dagger_\alpha$ is the adjoint operator to $\KK_\alpha$ and where $\frac{\delta \Xi}{\delta X^\alpha}\Big|_{X^\alpha = \KK_\alpha \frac{\delta S}{\delta \xx}} = J^\alpha$ is the thermodynamic flux corresponding to the force $X^\alpha$.

Evolution of constraint $c^\alpha(\xx)$ is then given by
\begin{equation}
    \dot{c}^\alpha = \left\langle d c^\alpha, \dot{\xx}\right\rangle
    = \left\langle d c^\alpha, \sum_\beta \KK^\dagger_\beta \frac{\delta \Xi}{\delta X^\beta}\Big|_{X^\beta = \KK_\beta \frac{\delta S}{\delta \xx}}\right\rangle
    = \sum_\beta \left\langle \KK_\beta d c^\alpha, \frac{\delta \Xi}{\delta X^\beta}\Big|_{X^\beta = \KK_\beta \frac{\delta S}{\delta \xx}}\right\rangle = 0,
\end{equation}
since $\KK_\beta d c^\alpha = 0$ by construction. 

For instance, in the case of energy conservation, $\xx=e(\rr)$, we have $\KK dE = 0$ with $\KK=\nabla$ and $dE = 1$. In the case of chemical reactions, $\xx = (n_1, n_2, \dots, n_N)$ are the number densities of $N$ chemical species and the dynamics is constrained by the admissible chemical reactions defined by the stoichiometric matrix $\nu^\alpha_i$, where $\alpha$ runs over all reactions and $i$ runs over all species. The constraints represent the conservation of atoms in the chemical reactions and have the form $C = \sum_i c_i n_i$. The operator $\KK$ is then given by $\KK_\alpha \xx^* = \sum_i \nu^\alpha_i x^*_i$, from which it follows that $\KK_\alpha dC = \sum_i \nu^\alpha_i c_i = 0$ since the number of atoms is conserved in each reaction. 

\subsection{GENERIC framework}
Gradient dynamics generates the dissipative part of the GENERIC framework \cite{grcontmath,grpla84,og,go,hco,pkg,be}
\begin{equation}\label{eq.generic}
\dot{\xx} = \{\xx,E\} + \frac{\delta \Xi}{\delta \xx^*}\Big|_{\xx^*=S_{\xx}}
\end{equation}
where $\{\bullet,\bullet\}$ is a Poisson bracket generating the reversible part of the evolution. Poisson brackets express the kinematics of the state variables $\xx$. Their specific forms in kinetic theory, continuum mechanics, complex fluids, plasma physics, classical mechanics, general relativity, elasticity, electrodynamics, and other parts of physics can be found, for example, in \cite{goldstein,gaybalmaz-ratiu,dv,pkg}.

Within GENERIC, the entropy is required to be a Casimir of the Poisson bracket, $\{S,A\}=0$ for any functional $A(\xx)$, which means that its evolution is not affected by the Hamiltonian part (Poisson bracket), and the second law of thermodynamics is satisfied due to the dissipative part. With the degeneracy condition on the dissipation potential given above \eqref{eq.Xi.deg}, energy is conserved, $\dot{E}=0$.

If the Poisson bracket is derived by reduction from a more microscopic level, which is the typical case, and if the energy, entropy, and dissipation potential are invariant with respect to the time-reversal transformation, GENERIC \eqref{eq.generic} automatically satisfies the Onsager-Casimir reciprocal relations \cite{hco,pkg}.

Other ways to derive the Poisson bracket are based on the underlying group and algebraic structure of the Poisson manifold, and they include the semidirect product construction \cite{marawe84}, the Lie-Poisson reduction \cite{gaybalmaz-ratiu}, and the method of matched pairs \cite{esen2016hamiltonian}. Also the Lagrange$\rightarrow$Euler transformation can be used to derive the Poisson bracket for continuum mechanics \cite{GodRom-elements,pepa-natural,romenski-viscous,shtc-generic}, which often leads to hyperbolic equations \cite{God-Siberian,ader-vis}. Finally, Poisson brackets can also be seen as classical versions of the quantum commutators \cite{dv}.

\subsection{Metriplectic dynamics}
In metriplectic dynamics \cite{mor,grcontmath, kauf,go, og}, the evolution is given by a sum of a Hamiltonian part and a gradient part,
\begin{equation}\label{eq.evo.metriplectic}
    \dot{\xx} = \{\xx,E\} +\MM(\xx)\cdot \frac{\delta S}{\delta \xx}
    \quad\text{or}\quad
    \dot{\xx} = \{\xx,E\} +[\xx, S]
\end{equation}
where $\MM$ is a symmetric and positive semi-definite operator that is degenerate in the direction of energy, $\MM\cdot \frac{\delta E}{\delta \xx} = 0$. The dissipative bracket of two functionals $A(\xx)$ and $B(\xx)$ is defined as $[A,B]=A_{x^i}M^{ij}B_{x^j}$, so the metriplectic evolution can either be written in terms of the dissipative operator $\MM$ or using the dissipative bracket.

Together with the skew-symmetry of the Poisson bracket, which implies that $\{E,E\}=0$, and with the property of entropy being a Casimir of the Poisson bracket, the degeneracy of the dissipative operator guarantees energy conservation, $\dot{E} = 0$. Positive semi-definiteness of that operator then implies the non-negativity of the entropy production, $\dot{S} \geq 0$.

How are metriplectic systems related to GENERIC \eqref{eq.generic}? GENERIC with a quadratic dissipation potential
\begin{equation}
    \Xi(\xx, \xx^*) = \frac{1}{2}\int d\rr\, \xx^*\cdot \MM(\xx)\cdot \xx^*
\end{equation}
can be seen as metriplectic dynamics \eqref{eq.evo.metriplectic}. However, not all metriplectic systems can be reformulated with a dissipation potential \cite{hutter2013,miroslav-guide}.

Moreover, $\MM$ is sometimes allowed to be non-symmetric \cite{hco}, which makes the system of equations incompatible with gradient dynamics.
Finally, metriplectic systems can be further generalized to the four-bracket formalism \cite{morrison4}.

\subsection{Entropy production maximization and Steepest Entropy Ascent}
Other frameworks that are related to gradient dynamics are variational frameworks where the entropy production is maximized subject to some constraints.
In the frameworks of Entropy Production Maximization \cite{ziegler.h:some,ziegler.h.wehrli.c:on,raja-sri,malek-maxwell}, the entropy production formula $J^\alpha X_\alpha$ is maximized subject to constraints given by a functional $\sigma(\mathbf{X})$. The functional actually represents the entropy production or a metric in the space of thermodynamic forces.

It was shown in \cite{adam-epm} that if the functional $\sigma(\mathbf{X})$ depends only on one force, or if it is $k$-homogeneous in all forces (for instance quadratic), the resulting force-flux relations are compatible with gradient dynamics \eqref{eq.GD}. However, in general, the resulting force-flux relations are not compatible with gradient dynamics.

A relation between the entropy production maximization and the principle of large deviations with a quadratic rate function for metriplectic systems has been shown in \cite{reina-zimmer}. 

A framework closely related to the entropy production maximization is the Steepest Entropy Ascent (SEA) \cite{sea}, which has been shown to be essentially equivalent with the GENERIC framework with a quadratic dissipation potential \cite{sea-generic}, that is with metriplectic systems. SEA represents a very general framework for dissipative evolution with many applications \cite{beretta-fourth}. 

In SEA, a strongly nondegenerate metric tensor $\gmetric$ on the state space $M$ is introduced, which provides a scalar product $(\mathbf{X},\mathbf{Y}) =\gmetric(\mathbf{X},\mathbf{Y})$ of two vectors $\mathbf{X}$ and $\mathbf{Y}$ in the tangent space of the state space. Then, in the presence of constraints $c_\alpha(\xx)$, the entropy gradient is first projected onto the subspace orthogonal to the gradients of the constraints, 
\begin{equation}
    ds^c = ds - \sum_\alpha \lambda^\alpha dc_\alpha,
\end{equation}
where the Lagrange multipliers $\lambda^\alpha$ are determined by the orthogonality condition $(ds^c, dc_\alpha) = 0$. 

The SEA evolution is then given by the sum of the reversible vector field $\XX_H$ and irreversible vector field $\YY_S$ where the latter is given by the steepest ascent of the entropy, $\YY_S = \frac{1}{\tau}\gmetric^{-1} ds^c$, where $\tau$ is a positive dimensional prefactor. The irreversible evolution then automatically respects the constraints, since $dc_\alpha(\YY_S) = (ds^c, dc_\alpha) = 0$. Moreover, the entropy production is non-negative, $\dot{S} = ds(\YY_S) = (ds^c, ds^c)/\tau \geq 0$. 

The SEA evolution can be also reformulated as a variational principle, where the entropy production is maximized subject to the constraint that the magnitude of the irreversible vector field $\YY_S$ is fixed, 
\begin{equation}
    \max_{\YY_S} ds^c(\YY_S) \quad\text{subject to}\quad \gmetric(\YY_S,\YY_S) = \text{const},
\end{equation}
where the dimensional prefactor $\tau$ then plays the role of the Lagrange multiplier of the constraint. In other words, the SEA framework can be seen as quadratic entropy production maximization that automatically respects the constraints. Alternatively, if the reversible vector field $\XX_H$ is Hamiltonian and does not affect the total entropy, the SEA framework can be seen as a metriplectic system (or GENERIC with quadratic dissipation potential).

Therefore, SEA has the same relation with GENERIC in its gradient form \eqref{eq.generic} as metriplectic systems \eqref{eq.evo.metriplectic}. Since the whole metric tensor $\gmetric$ must be supplied in SEA, it makes it more general than GENERIC in the gradient form, as not all metric tensors are Hessians of dissipation potentials. However, instead of one non-linear functional $\Xi$ (dissipation potential), SEA requires $n\cdot(n+1)/2$ functions (components of the metric tensor) to be supplied, where $n$ is the number of state variables. Moreover, the principle large deviations \cite{mielke-potential,hco-jnet2020-I,hco-jnet2020-II} leads to gradient dynamics with a generally non-quadratic dissipation potential.

\subsection{Onsager variational principle}
Another variational approach to dissipative dynamics is the Onsager variational principle \cite{onsager1930,doi-onsager}, which states that the evolution proceeds along the infimum of the following functional
\begin{equation}\label{eq.OVP}
    \Xi^*(\xx,\dot{\xx}) - \frac{\delta S}{\delta \xx}\cdot \dot{\xx}
\end{equation}
with respect to $\dot{\xx}$. The infimum is attained, assuming that $\Xi^*$ is sufficiently regular, convex, and coercive, at
\begin{equation}
    \frac{\delta \Xi^*}{\delta \dot{\xx}} = \frac{\delta S}{\delta \xx}.
\end{equation}

Equation \eqref{eq.OVP} can be seen as the Legendre transformation of dissipation potential $\Xi(\xx,\xx^*)$ with respect to $\xx^*$,
\begin{equation}
    \Xi^*(\xx,\dot{\xx}) = \sup_{\xx^*}\left(\xx^*\cdot \dot{\xx} - \Xi(\xx,\xx^*)\right),
\end{equation}
where the supremum is attained at the gradient dynamics \eqref{eq.GD}. Therefore, the Onsager variational principle is equivalent to the gradient dynamics.

\subsection{A variational principle for GENERIC}
There are several variational formulations of the GENERIC framework. One possibility is to use the formulation within contact geometry \cite{adv}. Another option is the Onsager variational principle for GENERIC \cite{omg1,omg2}. In the case of quadratic dissipation potential, also SEA provides a variational formulation of GENERIC \cite{sea-generic}. Finally, the principle of large deviations leads to a variational formulation of GENERIC \cite{mielke-potential,hco-jnet2020-I,hco-jnet2020-II,kraaj}.

Here, we present another variational formulation of GENERIC which is formulated for paths on the cotangent bundle $T^*M$ of the state space $M$. Assume that the Hamiltonian vector field $X_H:M\rightarrow TM$ is given by the Poisson bracket and energy, $X_H = \{\xx,H\}$. The cotangent bundle $T^*M$ is then parametrized by the state variables $\xx$ and their conjugate variables $\xx^*$. Then, stationary points of action 
\begin{equation}\label{eq.action}
    \int_0^T \left(-\Xi(\xx,\xx^*) + \xx^*(\dot{\xx}-X_H)\right) dt
\end{equation}
where $\Xi$ is a convex dissipation potential are given by Euler-Lagrange equations
\begin{subequations}\label{eq.lifted}
\begin{align}
    \label{eq.GENERIC}\dot{\xx} =& \frac{\partial \Xi}{\partial \xx^*} + X_H\\
    \label{eq.fiber}\dot{\xx}^* =& -\frac{\partial \Xi}{\partial \xx} - x^*_i \frac{\partial X_H^i}{\partial x^j},
\end{align}
where the first equation is the GENERIC equation \eqref{eq.generic} without the identification of $\xx^*$ with the gradient of entropy.
\end{subequations}

To make the GENERIC equation closed within the state variables $\xx$, the conjugate variables need to be identified with the derivatives of entropy, $\xx^*=dS$, forming a section $dS(\xx)$ of the cotangent bundle $T^*M$. The right-hand side of Equations \eqref{eq.lifted} represents the components of vector field $X_\Psi:T^*M\rightarrow TT^*M$ with $\Psi = -\Xi + \xx^*\dot{\xx}$,
\begin{equation}
    X_\Psi = \left(\frac{\partial \Xi}{\partial \xx^*} + X_H\right)\frac{\partial}{\partial\xx} - \left(\frac{\partial \Xi}{\partial \xx} + x^*_i \frac{\partial X_H^i}{\partial x^j}\right)\frac{\partial}{\partial x^*_j}.
\end{equation}
This vector field can be restricted to vector field $X^{dS}_\Psi:M\rightarrow TM$,
\begin{equation}
    X^{dS}_\Psi = T\pi_{M}\circ X_\Psi \circ dS = \left(X_H + \frac{\partial \Xi}{\partial \xx^*}\Big|_{\xx^*=dS}\right)\frac{\partial}{\partial\xx},
\end{equation}
through the commutative diagram
\begin{equation}
    \begin{array}{ccc}
        T^*M & \stackrel{X_\Psi}{\longrightarrow} & TT^*M\\
        \uparrow dS & & \downarrow T\pi_{M}\\
        M & \stackrel{X^{dS}_\Psi}{\longrightarrow} & TM
    \end{array}
\end{equation}
where $\pi_M:T^*M\rightarrow M$ is the canonical projection. The GENERIC equation \eqref{eq.lifted} then reduces to
\begin{subequations}
\begin{align}
    \dot{\xx} =& X^{dS}_\Psi\\
    \dot{S} =& \underbrace{X_H dS}_{=0} + \frac{\partial \Xi}{\partial \xx^*}\Big|_{\xx^*=dS}\geq0,
\end{align}
where the second law of thermodynamics is visible, assuming convexity of $\Xi$ in $\xx^*$ and that $S$ is conserved by the Hamiltonian vector field $X_H$.
\end{subequations}

Moreover, if $d\Xi(\xx,dS(\xx))=0$, the evolution for $\xx^*$ along the fiber \eqref{eq.fiber} is compatible with the GENERIC evolution \eqref{eq.GENERIC} in the sense that
\begin{equation}
    \dot{x}^*_i = \frac{d}{dt}\frac{\delta S}{\delta x^i} = \frac{\delta^2 S}{\delta x^i \delta x^j}\dot{x}^j
\end{equation}
so that only Equation \eqref{eq.GENERIC} needs to be solved. The Hamilton-Jacobi condition $\Xi(\xx,dS(\xx))=const$ has been identified in \cite{omg1} and if it is not satisfied for a dissipation potential, a function of $\xx$ can be added to the dissipation potential so that the condition becomes valid.

Note that the integrand in action \eqref{eq.action} actually depends on $\dot{\xx}$, but it can be rewritten as $-\Xi(\xx,\xx^*)dt + \xx^* d\xx - \xx^* X_H dt$, so we do not need to explicitly include the tangent direction $\dot{\xx}$ into the underlying geometry. Alternatively, we could go fully to the contact geometry \cite{adv}.

\section{Rayleigh dissipation potential}\label{sec.ray}

So far, we have based our analysis of dissipation on the approach in the mesoscopic state space to the classical equilibrium thermodynamics. The potential generating the dissipation was the entropy. Now we turn our attention to the approach in the space of mesoscopic vector fields to mesoscopic time evolution involving less microscopic details and more overall features. The new potential generating dissipation in such approach will be the \textit{Rayleigh dissipation potential}, that can also be called a \textit{rate entropy} since it is a real valued function of vector fields (in other words, a function of "rate state variables").

\subsection{Variational formulation}
Mechanical evolution of a physical system (described by state variables $\qq$) is obtained from the principle of stationary action, where the Lagrangian depends on $q$ and $\dot{q}$. The Rayleigh dissipation function $R(\dot{q}^i)$ then generates a force in the Euler-Lagrange equations \cite{Rayleigh,landau5},
\begin{equation}\label{eq.ray}
    \frac{d}{dt}\frac{\partial L}{\partial \dot{q}^i} - \frac{\partial L}{\partial q^i} = -\frac{\partial R}{\partial \dot{q}^i}.
\end{equation}
Potential $R$ is required to be positive and convex in $\dot{q}^i$ to guarantee the dissipation of energy,
\begin{equation}
    \dot{E} = -\frac{\partial R}{\partial \dot{q}^i}\dot{q}^i \leq 0.
\end{equation}
Equation \eqref{eq.ray} can be seen as a result of the variation of the mechanical action subject to constraints, as in \cite{bloch-krishnaprasad-marsden1996}, as we show later in Sec. \ref{sec.dalembert.ray}, where entropy is also added.

\subsection{Generalized formulation with Hamiltonian mechanics}
When the system has state variables $\xx = (\qq,s)$, where $s$ is the volumetric entropy density, the Rayleigh dissipation potential can be used to add dissipation to a Hamiltonian system (for instance, in fluid mechanics or complex fluids). The potential is then written in the form
\begin{equation}
    \RRR(\xx, \xx^\dagger) = \int d\rr \, \rrho(\xx, \xx^\dagger),
\end{equation}
which depends on the state variables $\xx$ and energetic conjugates $\xx^\dagger$ (to be identified with derivatives of energy $\xx^\dagger = \frac{\delta E}{\delta \xx}$), see \cite{dv}. The density of the dissipation potential is denoted by $\rrho$, and it can depend also on the spatial gradients of the conjugate fields. The time evolution is then given by
\begin{align}
    \partial_t \qq =& \{\qq,E\} -\frac{\delta \RRR}{\delta \qq^\dagger} = -\frac{\partial \rrho}{\partial \qq^\dagger} + \nabla\cdot \frac{\partial \rrho}{\partial \nabla \qq^\dagger}\\
    \partial_t s =& \{s,E\} -\frac{\delta \RRR}{\delta s^\dagger} + \frac{1}{s^\dagger}\left(\xx^\dagger\cdot \frac{\partial \rrho}{\partial \xx^\dagger}
    +\nabla\xx^\dagger \cdot\frac{\partial \rrho}{\partial \nabla \xx^\dagger}\right)\nonumber\\
    =& -\frac{\partial \rrho}{\partial s^\dagger} + \nabla\cdot\frac{\partial \rrho}{\partial \nabla s^\dagger} +  \frac{1}{s^\dagger}\left(\xx^\dagger\cdot \frac{\partial \rrho}{\partial \xx^\dagger}
    +\nabla\xx^\dagger \cdot\frac{\partial \rrho}{\partial \nabla \xx^\dagger}\right),
\end{align}
see \cite{dv}.

For energy conservation, it is required that $\rrho$ does not explicitly depend on $s^\dagger$, as such a dependence could be used for instance to describe the radiative heat transfer with the environment.
Energy is then conserved, as the evolution of the volumetric energy density $e(\xx)$ is of the divergence form,
\begin{equation}
    \partial_t e = -\nabla\cdot \JJ_e -\nabla\cdot\left(x^\dagger_\alpha \frac{\partial \rrho}{\partial \nabla x^\dagger_\alpha}\right),
\end{equation}
where $\JJ_e$ is the total energy flux generated by the Hamiltonian evolution (obtained from the evolution equations by the chain rule). Note, however, that in order to guarantee the chain rule, all fields and potentials must be sufficiently regular \cite{evans,roubicek}.

When the potential $\rrho$ is convex in both $\xx^\dagger$ and $\nabla \xx^\dagger$, the entropy production is non-negative,
\begin{equation}
    \partial_t s = -\nabla\cdot\JJ_s -\nabla\cdot\frac{\partial \rrho}{\partial \nabla s^\dagger} +  \underbrace{\left(\xx^\dagger\cdot \frac{\partial \rrho}{\partial \xx^\dagger}
    +\nabla\xx^\dagger \cdot\frac{\partial \rrho}{\partial \nabla \xx^\dagger}\right)}_{\geq 0},
\end{equation}
where $\JJ_s$ is the entropy flux generated by the Hamiltonian evolution. 
Both the first and second laws of thermodynamics are satisfied.

For instance, state variables for a fluid can be chosen as $\xx = (\rho,\mm,s)$, where $\mm$ is the momentum density, and the dissipation potential density is for a Newtonian fluid
\begin{equation}
    \rrho = \frac{1}{2}\mu \nabla \mm^\dagger : \nabla\mm^\dagger + \frac{1}{2}\lambda \nabla s^\dagger \cdot \nabla s^\dagger,
\end{equation}
where $\mu$ is the dynamic viscosity and $\lambda$ is a coefficient proportional to heat conductivity. The evolution equations then become
\begin{align}
    \partial_t \rho =& \{\rho,E\}\\
    \partial_t \mm =& \{\mm,E\} -\nabla\cdot(\mu \mm^\dagger) = -\nabla\cdot(\mu \vv),\\
    \partial_t s =& \{s,E\} + \nabla\cdot(\lambda \nabla s^\dagger) + \frac{1}{s^\dagger}\left(\nabla\mm^\dagger : \mu \nabla \mm^\dagger + \nabla s^\dagger \cdot \lambda \nabla s^\dagger\right)\nonumber\\
    =& -\nabla\cdot(-\lambda \nabla T) + \frac{1}{T}\left(\mu \nabla \vv : \nabla \vv + \lambda \nabla T : \nabla T\right),
\end{align}
where $\{\bullet,\bullet\}$ stands for the Poisson bracket for fluid mechanics \cite{arnold,dv,pkg}, $\vv = \mm^\dagger = E_\mm$ is the velocity field and $s^\dagger = E_s$ is the temperature. The evolution equations are the compressible Navier-Stokes-Fourier equations together with the heat equation with the viscous heating term.

\subsection{Gradient formulation}
The Rayleigh dissipation potential can be cast into the gradient dynamics framework by introducing dissipation potential
\begin{equation}
    \Xi(\xx^*, \xx^\dagger) = \int d\rr\, \xi\left(s^\dagger,\tqqs, \tss, \gtqq, \gtss\right)
    \quad\text{with}\quad
    \tqqs = \qq^* - \frac{\qq^\dagger}{s^\dagger},
    \tss = \frac{s^*}{s^\dagger},
    \gtqq = \nabla \qq^* - \frac{s^*}{s^\dagger}\nabla \qq^\dagger,
    \gtss = \nabla \frac{s^*}{s^\dagger},
\end{equation}
see \cite{pof2021}.
The dissipative evolution is then
\begin{subequations}\label{eq.R.evo.xi}
    \begin{align}
        \dot{\qq} =& \frac{\delta \Xi}{\delta \qq^*} = \frac{\partial \xi}{\partial \tqqs} - \nabla\cdot\frac{\partial \xi}{\partial \gtqq},\\
        \dot{s} =& \frac{\delta \Xi}{\delta s^*} = \frac{\partial \xi}{\partial \tss}\frac{1}{s^\dagger} - \frac{1}{s^\dagger}\frac{\partial \xi}{\partial \gtqq}\cdot\nabla \qq^\dagger - \frac{1}{s^\dagger} \nabla\cdot \frac{\partial \xi}{\partial \gtss}.
    \end{align}
\end{subequations}

On the Gibbs-Legendre (GL) manifold, both the entropic and energetic conjugate variables are identified with the corresponding derivatives of entropy and energy, $\xx^* = \frac{\delta S}{\delta \xx}$ and $\xx^\dagger = \frac{\delta E}{\delta \xx}$, which implies that $\qq^* = 0$, $s^* = 1$, $\qq^\dagger = \frac{\delta E}{\delta \qq}$, and $s^\dagger = T$. The evolution of energy density $e(\qq,s)$ is then
\begin{equation}
    \partial_t e = \frac{\partial e}{\partial \qq}\cdot \dot{\qq} + \frac{\partial e}{\partial s}\dot{s} = -\nabla\cdot\left(\frac{\partial e}{\partial \qq}\frac{\partial \xi}{\partial \gtqq} + \frac{\partial \xi}{\partial \gtss}\right)
    + \underbrace{\left(\qq^\dagger\cdot \frac{\partial \xi}{\partial \tqqs} + \frac{\partial \xi}{\partial \tss}\right)}_{=0 \text{ by construction}},
\end{equation}
where the condition
\begin{equation}
\frac{\qq^\dagger}{s^\dagger}\cdot \frac{\partial \xi}{\partial \tqqs} + \frac{s^*}{s^\dagger}\frac{\partial \xi}{\partial \tss} = 0
\end{equation}
is required for the energy conservation. This condition is analogous to condition \eqref{eq.Xi.deg} within gradient dynamics.

Moreover, on the GL manifold, it holds that $\tqqs = -\frac{1}{T}\frac{\delta E}{\delta \qq}$, $\tss = \frac{1}{T}$, $\gtqq = -\frac{1}{T}\nabla \frac{\delta E}{\delta \qq}$, and $\gtss = \nabla \frac{1}{T}$, and the dissipation potential density then becomes
\begin{equation}
    \xi|_{GL}
    =\xi\left(T, -\frac{E_\qq}{T}, \frac{1}{T}, -\frac{1}{T}\nabla E_\qq, \nabla \frac{1}{T}\right)
    \stackrel{def}{=} \xi^{GL}\left(T, \frac{E_\qq}{T}, \frac{1}{T}, \frac{1}{T}\nabla E_\qq, \nabla \frac{1}{T}\right).
\end{equation}
Equations \eqref{eq.R.evo.xi} then become
\begin{subequations}\label{eq.R.evo.xi.GL}
\begin{align}
\dot{\qq}|_{GL} =& -\frac{\partial \xi^{GL}}{\partial E_\qq/T} +\nabla\frac{\partial \xi^{GL}}{\partial \frac{1}{T} \nabla E_\qq},\\
\dot{s}|_{GL} =& -\nabla\cdot\left(\frac{1}{T} \frac{\partial \xi^{GL}}{\partial \nabla \frac{1}{T}}\right)
\underbrace{
+\frac{E_\qq}{T}\cdot \frac{\partial \xi^{GL}}{\partial E_\qq/T}
+\nabla\frac{1}{T} \cdot \frac{\partial \xi^{GL}}{\partial \nabla \frac{1}{T}}
+\frac{\nabla E_\qq}{T} \cdot \frac{\partial \xi^{GL}}{\partial \frac{1}{T}\nabla E_\qq}}_{\geq 0},
\end{align}
where the non-negativity of the entropy production follows from the convexity (or rather the radial monotonicity) of $\xi^{GL}$ in its four last arguments.
\end{subequations}
The Rayleigh dissipation potential in the gradient formulation thus leads to the same evolution equations as the classical formulation.

Similarly to gradient dynamics, the Rayleigh dissipation potential provides a framework where energy is dissipated into the internal energy by raising the temperature \cite{Rayleigh,landau-ginzburg,bloch-dissipation}. It is less general than the gradient dynamics framework, as it requires the volumetric entropy density among the state variables. However, it is more general than classical irreversible thermodynamics, as it does not require conserved state variables and allows for non-linear constitutive relations.

\subsection{Single generator formalism}
In the Single Generator Formalism \cite{be}, the evolution of state variables $\xx$ is given by the sum of their Hamiltonian evolution and dissipative evolution, where the latter is generated by a dissipative bracket and energy $E$. The bracket has the form
\begin{equation}
[F,G] = \int d\rr\, \xi\left(L\left(F_\qq,\nabla F_\xx\right);G_\xx,\nabla G_\xx\right)
-\int d\rr\, F_s \xi\left(L\left(G_\qq,\nabla G_\xx\right);G_\xx,\nabla G_\xx\right)
\end{equation}
where $L$ is a linear operator and $\xi$ turns the derivatives of functionals into a density. The evolution equations obtained from this bracket, when $G$ is taken as the energy $E$, are identical to those obtained from the Rayleigh dissipation potential in the gradient formulation \eqref{eq.R.evo.xi}, when $\rrho=-\xi$. Therefore, the dissipation within the Single Generator Formalism is equivalent to Hamiltonian mechanics with a quadratic Rayleigh dissipation potential.

\subsection{Extended GENERIC}
An alternative route towards the Rayleigh dissipation potential is to start from the GENERIC equation and to extend it by adding additional state variables $\ww$ that evolve faster than the original state variables $\xx$. The fast evolution of $\ww$ then leads to the Rayleigh dissipation potential for the slow evolution of $\xx$.

We begin by extending the GENERIC time evolution equation with state variables $\xx$ to another GENERIC time evolution equation with state variables $(\xx,\ww)$
\begin{equation}\label{Eq1}
\left(\begin{array}{cc}\dot{\xx}\\ \dot{\ww}\end{array}\right)=\left(\begin{array}{cc}\LL&\KK^\dagger\\-\KK&0\end{array}\right)\left(\begin{array}{cc}
\xx^*\\ \ww^*\end{array}\right) - \left(\begin{array}{cc}0\\ \Xi_{\ww^*}\end{array}\right)
\end{equation}
where $\LL$ is the Poisson bivector expressing the Hamiltonian kinematics of $\xx$ (i.e., $\{A,B\}=\langle A_{\xx},\LL B_{\xx}\rangle)$, $\KK$ is a linear operator, $\KK^\dagger$ is its adjoint, $(\xx^*,\ww^*)$ are conjugate state variables, and $\Xi(x,X,\ww^*)$ is a dissipation potential. We assume that $\Xi$ is independent of $\ww$ and, as a function of $\ww^*$, it satisfies all the properties required of the dissipation potential \cite{pkg}. Equation (\ref{Eq1}) is indeed a particular realization of the GENERIC equation, except for the Jacobi identity, which is not necessarily required in the presence of dissipation. If external forces are absent, then there exists a thermodynamic potential $\Phi(\xx,\ww)$ and  $\dot{\Phi}\leq 0$ provided $\xx^*=\Phi_{\xx}$ and $\ww^*=\Phi_{\ww}$, typically $\Phi(\xx) = -S(\xx,\ww) + \frac{1}{T_0}E(\xx,\ww)$ where $T_0$ is a fixed parameter (the equilibrium temperature).

We note that the second equation in (\ref{Eq1}) governs the time evolution of $\ww^*$ provided we transform its left-hand side into $\dot{\ww}^*$. We now assume that $\ww$ evolves faster than $\xx$, so in the second equation in (\ref{Eq1}) $\xx$ is treated as a fixed parameter for the moment.
The second equation in (\ref{Eq1}) governing the fast time evolution of $\ww^*$ becomes
\begin{equation}\label{Eq2}
\dot{\ww}^*=-\Phi_{\ww\ww}\RRR_{\ww^*}
\end{equation}
where
\begin{equation}\label{Eq3}
\RRR(x, \ww^*,X)=\Xi(\ww^*,X)+\langle \KK \xx^*,\ww^* \rangle
\end{equation}
is the rate thermodynamic potential called the Rayleigh potential and $X=\KK\xx^*$ a thermodynamic force. Consequently, $\dot{\RRR}=-\langle\RRR_{\ww^*},\Phi_{\ww\ww}\RRR_{\ww^*}\rangle\leq 0$, as the Hessian $\Phi_{\ww\ww}$ is a positive definite operator, assuming sufficient regularity and locality. As for solutions of (\ref{Eq2}), we note that $\RRR$ decreases with $t\rightarrow \infty$ and potential $\RRR$ then plays the role of the Lyapunov function (or S-function).

When we construct the dual dissipation potential $\Xi^*(\xx,X,J_\ww)$ by the Legendre transformation $\Xi^* = -\Xi + \langle J_\ww, \ww^*\rangle$, $J_\ww = \Xi_{\ww^*}$, $\ww^* = \Xi^*_{J_\ww}$, the quasi-equilibrium solution of (\ref{Eq1}) becomes $\ww^* = \Xi^*_{J_\ww}\Big|_{J_\ww = -X}$, and the remaining equation for the slow evolution of $\xx$ becomes
\begin{equation}\label{Eq5}
    \dot{\xx}=\LL\xx^*+\KK^{\dag}\Xi^*_{J_\ww}\Big|_{J_\ww = -X},
\end{equation}
which has the GENERIC form for the $\xx$ variable. In particular, we see that $\dot{\Phi}(\xx)=-\langle X,\Xi^*_X \rangle\leq 0$. More detailed formulation of the extended GENERIC equation can be found in \cite{mg2024, atesli2025}.

\section{Dissipative d'Alembert variational principle}\label{sec.var}
\subsection{General formulation}
The d'Alembert variational principle with dissipation was formulated in \cite{fgb19-variational-thermodynamics} as a generalization of the classical d'Alembert principle \cite{bloch-krishnaprasad-marsden1996}. The action
\begin{equation}
    \int_0^T L(\qq,\dot{\qq},S) dt + \int_0^T \langle \FF^{\text{ext}}, \delta \qq\rangle dt,
\end{equation}
where $\FF^{\text{ext}}$ is an external force,
which is varied subject to the constraints
\begin{equation}
\frac{\partial L}{\partial S} \delta S = \langle \FF^{\text{diss}}, \delta \qq\rangle
\quad\text{and}\quad
\frac{\partial L}{\partial S} \dot{S} = \langle \FF^{\text{diss}}, \dot{\qq}\rangle,
\end{equation}
where $\FF^{\text{diss}}(\qq,\dot{\qq},S)$ is a dissipative force. The resulting equations of motion are
\begin{subequations}\label{eq.dalembert.general}
\begin{align}\label{eq.dalembert}
    \frac{d}{dt}\left(\frac{\partial L}{\partial \dot{\qq}}\right) - \frac{\partial L}{\partial \qq} =& \FF^{\text{ext}} + \FF^{\text{diss}}\\
    \dot{S} =& \frac{1}{L_S}\langle \FF^{\text{diss}}, \dot{\qq}\rangle,
\end{align}
\end{subequations}
where the first equation is the classical d'Alembert principle with dissipation \cite{landau5}. For the derivative of the Lagrangian with respect to entropy, it holds that $L_S = -H_S = -T < 0$, and the second law of thermodynamics is thus satisfied ($\dot{S}\geq 0$), for instance, when $F_i^{\text{diss}} = -\lambda_{ij}\dot{q}^j$ with $\lambda_{ij}$ being a positive definite matrix.

\subsection{Rayleigh-d'Alembert variational principle}\label{sec.dalembert.ray}
Let us now restrict the general formulation of the dissipative d'Alembert variational principle to the case when the dissipative forces are generated by a Rayleigh dissipation potential. The action
\begin{subequations}
\begin{equation}
    \int_0^T L(\qq,\dot{\qq},S) dt
\end{equation}
is then varied subject to the constraints
\begin{equation}
\frac{\partial L}{\partial S} \delta S = -\frac{\partial R}{\partial \dot{\qq}}\delta{\qq}
\quad\text{and}\quad
\frac{\partial L}{\partial S} \dot{S} = -\frac{\partial R}{\partial \dot{\qq}}\dot{\qq},
\end{equation}
where $R(\dot{\qq})$ is a convex dissipation potential. This is a special case of the principle from \cite{fgb19-variational-thermodynamics}.
\end{subequations}
Then, the equations of motion are
\begin{subequations}\label{eq.dalembert.ray}
\begin{align}
    \frac{d}{dt}\left(\frac{\partial L}{\partial \dot{\qq}}\right) - \frac{\partial L}{\partial \qq} =& -\frac{\partial R}{\partial \dot{\qq}}\\
    \dot{S} =& -\frac{1}{L_S}\frac{\partial R}{\partial \dot{\qq}}\dot{\qq},
\end{align}
\end{subequations}
where the first equation is the Landau evolution equation 121.8 \cite{landau5} and the second equation stands for the second law of thermodynamics. As the derivative $L_S$ is negative, the entropy production is positive. In the case of quadratic dissipation potential, $R = \frac{1}{2}\lambda_{ij}\dot{q}^i\dot{q}^j$ with $\lambda_{ij}$ being a positive definite matrix, the equations reduce to the d'Alembert equations \eqref{eq.dalembert.general} with dissipative forces $F_i^{\text{diss}} = -\lambda_{ij}\dot{q}^j$.

\subsubsection{Hamiltonian picture}
Equations \eqref{eq.dalembert.ray} can be also transformed to their Hamiltonian form, where the Hamiltonian is the Legendre transform of the Lagrangian,
\begin{subequations}
\begin{align}
    \dot{\qq} =& \frac{\partial H}{\partial \pp}\\
    \dot{\pp} =& -\frac{\partial H}{\partial \qq} - \frac{\partial R}{\partial H_\pp}\\
    \dot{S} =& \frac{1}{H_S}\frac{\partial R}{\partial H_\pp}H_\pp,
\end{align}
where $H_S=-L_S$ was used. The Hamiltonian is conserved, $\dot{H}=0$, as follows from the chain rule, and the entropy production is positive, $\dot{S}>0$. In the context of the d'Alembert principle, the Hamiltonian picture was formulated in \cite{yoshimura-hamiltonian}.
\end{subequations}

Note, however, that the Lagrangian may be singular, and thus the Legendre transformation does not need to be diffeomorphic. In that case, one may use the framework of Tulczyjew triples \cite{tulczyjew-legendre}, using Lagrangian submanifolds. Other possibility is the use of the slope transformation \cite{dorst1993,dorst1994,nonconvex}, where the Legendre-transformed potentials become graphs (or multivalued functions).

Near thermodynamic equilibrium, the dissipation potential can be approximated by a quadratic functional,
\begin{equation}
    R = \frac{1}{2}\Lambda_{ij}H_{p_i}H_{p_j},
\end{equation}
where $\Lambda_{ij}$ is a positive definite matrix and the matrix can be taken as symmetric without any loss of generality.
When the dissipation potential is invariant with respect to the time-reversal transformation, the Onsager-Casimir reciprocal relations \cite{onsager1930,onsager1931,casimir1945} are then automatically valid, as the Hamiltonian part satisfies them automatically (see \cite{pkg}) while the dissipative part gives
\begin{subequations}
\begin{align}
    \dot{q}^i =& H_{p_i}\\
    \dot{p}_i =& -H_{q^i} -\Lambda_{ij}H_{p_j}
\end{align}
with $\Lambda_{ij}=\Lambda_{ji}$.
\end{subequations}

\subsubsection{Reduced dynamics}
The Hamiltonian version of the d'Alembert principle can now be used to reduce the system of equations, assuming that a momentum variable relaxes quickly to an equilibrium value where $\dot{\pp}\approx 0$. The equation for $\pp$ then simplifies to 
\begin{equation}
    0 = -H_\qq - \frac{\partial R}{\partial H_\pp}.
\end{equation}
This can, however, be seen as a Legendre transformation defining the dependence $\tilde{H}_\pp(H_\qq)$ and the dual dissipation potential $R^*(H_\qq)$,
\begin{equation}
    R^*(H_\qq) = -H_\qq \tilde{H}_\pp(H_\qq) - R(\tilde{H}_\pp(H_\qq)),
\end{equation}
derivative of which is
\begin{equation}
    \frac{\partial R^*}{\partial H_\qq} = -\tilde{H}_\pp(H_\qq).
\end{equation}
The second derivative of $R^*$ is positive definite, as the Hessian of $R$ is positive definite (equal to the inverse of $\boldsymbol{\Lambda}$ in the quadratic case).
Then, the Hamiltonian equations reduce to
\begin{subequations}
\begin{align}
    \dot{\qq} =& -\frac{\partial R^*}{\partial H_\qq}\\
    \dot{S} =& \frac{1}{H_S}\frac{\partial R^*}{\partial H_\qq}H_\qq\geq 0.
\end{align}
Entropy production is still positive while keeping the Hamiltonian constant, $\dot{H}=0$.
\end{subequations}

\section{Dissipation constructed from the Hamiltonian structure}\label{sec.ham}
Another class of frameworks for dissipative evolution is constructed from the Poisson bracket and from the energy of a Hamiltonian system.

\subsection{Double bracket dissipation}
Double bracket dissipation \cite{brockett1988,bloch-dissipation,bloch-krishnaprasad-marsden1996} is a framework that adds to a Hamiltonian system dissipative terms. 

For a Lie-Poisson system, the Poisson bracket has the structure 
\begin{equation}
    \{F,G\} = [F_\xx, G_\xx]\cdot \xx = C_{ij}^k x_k F_{x_i}G_{x_j}
\end{equation}
where $[\mathbf{a},\mathbf{b}]^k=C_{ij}^k a^i b^j$ is a Lie bracket and $C_{ij}^k=-C_{ji}^k$ are the structure constants \cite{fecko}. The Hamiltonian evolution is then 
\begin{equation}
    (\dot{x}_i)_{rev} = \{x_i,H\} = C_{ij}^k x_k H_{x_j}.
\end{equation}
A Casimir invariant $S$ satisfies $\{S,F\}=0$ for all functionals $F$, which implies that $S_{x_i} C_{ij}^k x_k = 0$.

The double bracket dissipation is described by evolution equation
\begin{equation}
    \dot{x}_i = C_{ij}^k x_k H_{x_j} + \alpha C_{ij}^k x_k C_{jl}^m x_m H_{x_l},
\end{equation}
which for functionals of the state variables gives
\begin{align}
    \dot{F} =& F_{x_i}\dot{x}_i
    = F_{x_i} C_{ij}^k x_k H_{x_j} + \alpha F_{x_i} C_{ij}^k x_k C_{jl}^m x_m H_{x_l},\nonumber\\
    =&
    \{F,H\} + \alpha F_{x_i} C_{ij}^k x_k C_{jl}^m x_m H_{x_l}
\end{align}
where $\alpha>0$ is a constant that controls the strength of dissipation, $H$ is the Hamiltonian, and $H_\xx$ is the functional derivative of the Hamiltonian with respect to the state variables.

The Hamiltonian behaves as
\begin{equation}
    \dot{H} = \{H,H\} + \alpha H_{x_i} C_{ij}^k x_k C_{jl}^m x_m H_{x_l} 
    = 0 - \alpha \langle H_{x_i} C_{ij}^k x_k, H_{x_l} C_{lj}^m x_m\rangle \leq 0,
\end{equation}
where $\langle\bullet,\bullet\rangle$ is the inner product,
and Casimirs $S$ as
\begin{equation}
    \dot{S} = \{S,H\} + \alpha S_{x_i} C_{ij}^k x_k C_{jl}^m x_m H_{x_l} 
    = 0 + \alpha \left(S_{x_i} C_{ij}^k x_k\right) \left(H_{x_l} C_{lj}^m x_m\right) = 0,
\end{equation}
so they are conserved. Therefore, the double bracket dissipation dissipates energy while conserving Casimirs.

\subsection{Ehrenfest regularization}
Ehrenfest regularization \cite{ehre} is another framework that adds to a Hamiltonian system dissipative terms constructed from the Poisson bracket and energy. It is based on the method of Ehrenfest reduction \cite{gk-ehrenfest,gk}, where the Hamiltonian vector field is prolonged and then projected to a reduced state space. For a system with state variables $\xx$ and Poisson bracket $\{A,B\} = A_{x^i}L^{ij}B_{x^j}$ (which is a general form of a Poisson bracket, $\mathbf{L}$ being called the Poisson bivector), the Hamiltonian evolution is given by
\begin{equation}
    (\dot{\xx})_{rev} = \{\xx,H\} = L^{ij} H_{x^j}.
\end{equation}
Casimir invariants $S$ satisfy $\{S,F\}=0$ for all functionals $F$, which implies that $\mathbf{L}\cdot S_\xx = 0$.

The Ehrenfest regularization comes in two forms: (i) energetic Ehrenfest regularization (E-EhRe) that dissipates energy while conserving the Casimirs, and (ii) entropic Ehrenfest regularization (S-EhRe) that conserves energy while producing entropy (or Casimirs in general). 

\subsubsection{Energetic Ehrenfest regularization}
The evolution equations for E-EhRe are
\begin{equation}
    \dot{x}^i = \{x^i,H\} - \frac{\tau}{2} \underbrace{L^{ki}H_{x^k x^l} L^{lj}}_{=M^{ij}} H_{x^j}
\end{equation}
where $\tau>0$ is a constant that controls the strength of dissipation, and the operator $\mathbf{M}$ is positive semidefinite and symmetric by construction (assuming convexity of $H$). The evolution of energy is
\begin{equation}
    \dot{H} = \{H,H\} - \frac{\tau}{2} \left(L^{ki}H_{x^i}\right) H_{x^k x^l}\left( L^{lj} H_{x^j}\right) \leq 0, 
\end{equation}
as $\{H,H\} = 0$,
and the evolution of Casimirs $S$ is
\begin{equation}
    \dot{S} = \{S,H\} - \frac{\tau}{2} S_{x^i} L^{ki}H_{x^k x^l} L^{lj} H_{x^j} 
    = 0,
\end{equation}
since $S_{x^i} L^{ki} = 0$,
so Casimirs are conserved. This behavior is suitable for instance in dissipative rigid-body dynamics, 
\begin{equation}
    \dot{\mm} = \mm\times \oomega + \frac{\tau}{2} \mm\times \left( \mathbb{I}^{-1} (\mm\times \oomega) \right)
\end{equation}
where $\mm$ is the angular momentum in a reference frame fixed with the body, $\mathbb{I}^{-1}=d^2H$ is the inverse of the inertia tensor, and $\oomega = \mathbb{I}^{-1} \mm$ is the angular velocity. Energy then decreases in time while keeping the Casimirs (the magnitude of angular momentum). 

\subsubsection{Entropic Ehrenfest regularization}
The evolution equations for S-EhRe are
\begin{equation}\label{eq.ehre.S}
    \dot{x}^i = \{x^i,H\} +\underbrace{L^{ij}_k L^{kl}H_{x^l}}_{=N^{ij}} H_{x^j}
\end{equation}
where the operator $\mathbf{N}$ is antisymmetric by construction (assuming smoothness of $L^{ij}$) and where $L^{ij}_k = \frac{\delta L^{ij}}{\delta x^k}$ stands for the functional derivative of the Poisson bivector.
The evolution of energy is
\begin{equation}
    \dot{H} = \{H,H\} + H_{x^i}L^{ij}_k L^{kl}H_{x^l} H_{x^j} 
    = 0,
\end{equation}
and energy is thus conserved due the skew-symmetry of $\mathbf{N}$.
The evolution of Casimirs is
\begin{equation}
    \dot{S} = \{S,H\} + S_{x^i} L^{ij}_k L^{kl}H_{x^l} H_{x^j} 
    = 0 + \frac{\delta}{\delta x^k}\underbrace{\left(S_{x^i} L^{ij}\right)}_{=0} L^{kl}H_{x^l} H_{x^j}
    -S_{x^i x^k} L^{ij} L^{kl}H_{x^l} H_{x^j}
    \geq 0,
\end{equation}
so entropy, or any concave Casimir, is being produced (convex Casimirs are being reduced). This behavior can be used to generate dissipation in kinetic theory, 
\begin{equation}
    \partial_t f = -\frac{p_i}{m}\frac{\partial f}{\partial r^i} + \frac{\tau}{2} \frac{p_k}{m}\frac{p_l}{m}\frac{\partial^2 f}{\partial r^k \partial r^l},
\end{equation}
or fluid mechanics, where similar terms appear as in \cite{svard}, where Laplacians of the state variables are added to all evolution equations in fluid mechanics.

\section{Comparison of the Port-Hamiltonian and the Extended GENERIC views of Thermodynamics of Externally Driven Systems}\label{sec.ph}

When carrying thermodynamics from externally unforced to externally driven systems the state variables shift from the base space to its tangent and cotangent spaces. This result has appeared in the context of the GENERIC formulation of thermodynamics, and it has also appeared in its Port-Hamiltonian (PH) formulation \cite{van2004port}. In the former it is because thermodynamics is associated with dynamics describing the approach to fixed points. In externally unforced systems, the fixed points are thermodynamic equilibrium states in the base space. In externally driven systems, the fixed points are vector fields generating the time evolution in reduced dynamical theories, involving less detail. From the PH viewpoint, the necessity to turn attention to tangent and cotangent spaces has appeared in the attempt to bring the power of thermodynamics to the control theory \cite{maschke-schaft,schaft-survey}.

We hope that the comparison of the Extended GENERIC (\ref{Eq1}) with the PH thermodynamics that we formulate below will help to combine and enlarge the pools of insights that emerged independently in both GENERIC and PH views of thermodynamics.

Vector space
$W$ with elements $w=(\xi,w_R,w_{ext})\in W$ plays the role of state variables in PH-thermodynamics. The first vector is $\xi\in T_xM$, where $M$ is the state space GENERIC-thermodynamics; $x\in M$. The vector $w_R$ is called a resistive state variable and $x_{ext}$ is a vector expressing an external influence (like control and boundary). 
The vector space $W^*$ with elements $w^*=(\xi^*,w^*_B,w^*_{ext})\in W^*$ is dual to $W$. The element $\xi^*$ is physically interpreted as gradient of the energy $H(x)$, i.e. $\xi^* =H_x$. To each pair $(w,w^*)$, we associate $\langle w^*,w\rangle\in \mathbb{R}$. 

Arguments developed in the control theory and thermodynamics \cite{maschke-schaft} led to the following result. Processes that are compatible with the Hamiltonian dynamics and thermodynamics are restricted to a submanifold $\mathcal{D}\subset W\oplus W^*$ determined by (see e.g. Eq.(30) in (\cite{schaft-survey}))
\begin{equation}\label{PH1}
\left(\begin{array}{ccc}\dot{x}\\w_R\\w_{ext}\end{array}\right)=\left(\begin{array}{ccc}\mathbb{L}&\KK_{R}(x)&\KK_{ext}(x)\\-\KK_R^\dagger(x)&0&0\\
-\KK_{ext}^\dagger(x)&0&0\end{array}\right)\left(\begin{array}{ccc}H_x\\w_R^*\\w_{ext}^*\end{array}\right)
\end{equation}
and
\begin{equation}\label{PH2}
R(w_R,w^*_R)=0\,\,\text{guaranteeing}\,\,\langle w^*_R,w_R\rangle\leq 0\,\forall\,\,(w^*_R,w_R)
\end{equation}
where $\KK_R(x)$ and $\KK_{ext}(x)$ are linear operators, $R$ is in general a nonlinear function of $(w_R,w^*_R)$.

Now we compare (\ref{PH1}) with (\ref{Eq1}). With $(w_R,w_R^*,w_{ext},w_{ext}^*)$ absent in (\ref{PH1}) and $w$ absent in (\ref{Eq1}), the first equations in both (\ref{PH1}) and (\ref{Eq1}) are the same.
To compare the second equations in (\ref{PH1}) and (\ref{Eq1}) we interpret $w_R$ as an element $\dot{\zeta}_R$ of the tangent space $TN_R$ of the space $N_R$ with elements $\zeta_R \in N_R$. With $(w_{ext},w_{ext}^*)$ absent, the first two equations in (\ref{PH1}), describing now the time evolution in an extended state space $M\times N_R$,  become
\begin{equation}\label{PH3}
\left(\begin{array}{cc}\dot{x}\\ \dot{\zeta}_R\end{array}\right)=\left(\begin{array}{cc}\mathbb{L}&\KK_{R}(x)\\-\KK_R^\dagger(x)&0\end{array}\right)\left(
\begin{array}{cc}H_x\\w_R^*\end{array}\right).
\end{equation}
This reinterpretation of $w_R$ allows us to express the resistivity in a way that is physically more transparent  than the requirement (\ref{PH2}). 

We assume that the resistivity is generated in the time evolution of $\zeta_R$.
This assumption is expressed mathematically by modifying the second equation in (\ref{PH3}). We add to its right-hand side the dissipative term in the same way as we do in (\ref{Eq1}), and the modified second equation in (\ref{PH3}) becomes
\begin{equation}\label{PH4}
\dot{\zeta}_R=-\KK^T_RH_x+\Xi_{w^*}(w^*,x).
\end{equation}
If the dissipation is strong (i.e. $\Xi_{w^*}(w^*,x)$ is dominant on the right-hand side of (\ref{PH4})), then $\zeta_R$ evolves rapidly to
\begin{equation}\label{PH5}
\dot{\zeta}_R=0=-K^T_RH_x+\Xi_{w^*}(w^*,x).
\end{equation}
Summing up, we have replaced the two equations
\begin{eqnarray}\label{PH6}
w_R&=&-K_R^T(x)H_x(x)\nonumber \\
R(w_R,w^*_R)&=&0
\end{eqnarray}
that allow to eliminate the resistive state variables $(w_R,w^*_R)$ from the PH equations  (\ref{PH1}) with Eq.(\ref{PH5}).

If in the third equation in (\ref{PH1}) we interpret $w_{ext}$ as an element $\dot{\zeta}_{ext}$ of the tangent space $TN_{ext}$ of the space $N_{ext}$ with elements $\zeta_{ext}\in N_{ext}$ then the PH formulation (\ref{PH1}), (\ref{PH2}) of thermodynamics of driven systems becomes completely equivalent to the Extended GENERIC formulation (\ref{Eq1}) describing the time evolution in the state space $M\times N_R\times N_{ext}$.

In both the formulations of extensions of thermodynamics from externally unforced to externally driven systems, the extended state variables are rate state variables (i.e. elements of tangent and cotangent bundles). In the Extended GENERIC formulation they emerge as elements of tangent and cotangent spaces of an extended state space. From the physical point of view, the enlargement of the base space means that the external influences are internalized.
In the PH formulation the state space (the base space of the bundles) remains unchanged, only its tangent and cotangent spaces are enlarged.

This type of extension has also been investigated in the GENERIC formulation in \cite{loma2023} in the context of chemical kinetics. 
From the mathematical point of view, the extension in \cite{loma2023} involves the structure of Lie algebroids \cite{pradines1967}, in the PH extension it involves the Dirac structure \cite{courant-weinstein} (the submanifold $\mathcal{D}\subset W\oplus W^*$  specified explicitly in (\ref{PH1}) appears to be a Dirac structure on $W\oplus W^*$). We intend to explore both the physical and the mathematical aspects of these two views of extensions in the future.

\section{Contact structure geometry}\label{sec.contact}

We have seen that by adding dissipation to the microscopic Hamiltonian dynamics, we bring attention to overall features of the dynamics that are seen in our everyday observations of the behavior of macroscopic systems. The pure nondissipative Hamiltonian dynamics as well as the pure dissipative dynamics are two extreme idealizations of our experience. But the dynamics at these two extremes are, from the mathematical point of view,
understood best. In the former case the fundamental group of the dynamics is the group of canonical transformation (transformations preserving symplectic structure) and in the latter case it is the group of Legendre transformations (transformations arising in maximization of entropy in the presence of constraints).

What is the group, combining both of these two groups, that is the fundamental group of mesoscopic dissipative dynamics?
In order to answer this question, we turn to the strategy introduced by Lagrange \cite{lagrange} in his analysis of maximization of a potential in the presence of constraints. The first step is counterintuitive. We naturally expect that we find an insight needed for solution in a smaller and thus a simpler setting. Instead, Lagrange begins with enlarging the setting. He looks for maximization of a new enlarged potential that depends on extra variables called Lagrange multipliers. In such enlarge setting the desired insights and simplifications then emerge.

Following the Lagrange strategy, we begin the process of bringing micro and macro dynamics to a harmony on the side of macro (i.e., in classical equilibrium thermodynamics). We consider the conjugate state variables $\xx^*$ as quantities playing the role of Lagrange multipliers and then, together with the entropy $s$ that is the maximized potential, we adopt them as extra state variables. The additional new step now is to realize that such an enlarged space is naturally equipped with the contact structure defined by the 1-form $\eta=ds-\xx^*d\xx$. This 1-form also has, in the context of classical equilibrium thermodynamics, an important physical significance. It can be shown \cite{grmela1990, ent16, pkg, omg1,omg2} that
the GENERIC dynamics becomes, on the Legendre manifold (i.e., a submanifold in the enlarged space on which the 1-form $\eta=0$), dynamics preserving
the contact structure.

We have thus succeeded in putting dissipation addressing thermodynamics into harmony with Hamiltonian mechanics addressing the microscopic origin of macroscopic systems. The thermodynamic aspects (the fundamental thermodynamic relations on both the starting and target levels) are expressed in the geometry of the manifold (which is the Legendre manifold) on which the time evolution takes place and in the dynamics in the contact Hamiltonian generating the time evolution. The geometrical structure transforming gradient of the potential to vector field is the contact structure that is preserved in the GENERIC time evolution.

\section{Conclusion}
We have reviewed several geometric frameworks for dissipation in non-equilibrium thermodynamics, highlighting their key features and interconnections. Each framework offers unique advantages and is suited for different types of problems in thermodynamics.

In particular, the gradient dynamics framework provides a versatile and general approach to modeling dissipative processes, and it is supported by statistical physics. The Rayleigh dissipation potential offers a more specific formulation that is particularly useful when the volumetric entropy density is among the state variables, that is, in continuum thermodynamics. Moreover, the formulation with the Rayleigh dissipation potential can be seen as a special case of gradient dynamics and of the GENERIC framework. The GENERIC framework, where Hamiltonian mechanics is combined with gradient dynamics, can also be formulated as a variational principle.

The dissipative d'Alembert principle provides a variational formulation for non-equilibrium thermodynamics and can be brought close to the framework of the Rayleigh dissipation potential and the GENERIC framework. The conjugate Hamiltonian formulation can then be used to reduce fast-relaxing momentum variables. 

Finally, dissipative dynamics can also be constructed from the underlying Poisson bracket. The dissipative dynamics can then either produce Casimirs (entropy) and keep the energy constant or reduce the energy while keeping the Casimirs.

In the future, we would like to explore how these dissipative frameworks are able to handle the problem of boundary conditions \cite{eldred-gb-brackets}.

\section*{Acknowledgment}
MP was supported by Czech Science Foundation, project 23-05736S.
MP is a member of the Nečas Center for Mathematical Modeling.
MP used Github Copilot for grammar cleaning and for minor code suggestions.
We are grateful to O{\u g}ul Esen and Fran{\c c}ois Gay-Balmaz for valuable discussions on differential geometry and its applications in non-equilibrium thermodynamics. We also thank Mark Peletier and Marcello Seri for discussing the statistical background of gradient dynamics.


\end{document}